\documentclass[aps,prl,twocolumn,showpacs,superscriptaddress,floatfix]{revtex4}
\usepackage{graphicx} 
\usepackage{bm}

\begin{document}

\title{Curie law, entropy excess, and superconductivity in heavy
  fermion metals and other strongly interacting Fermi liquids}

\author{V.~A.~Khodel}
\affiliation{Russian Research Centre Kurchatov Institute, Moscow,
  123182, Russia}

\author{M.~V.~Zverev}
\affiliation{Russian Research Centre Kurchatov
  Institute, Moscow, 123182, Russia}

\author{Victor M.~Yakovenko} \affiliation{Condensed Matter Theory
  Center and Center for Superconductivity Research, Department of
  Physics, University of Maryland, College Park, Maryland 20742-4111,
  USA}

\date{cond-mat/0508275 v.2, 10 December 2005, Phys. Rev. Lett.
  {\bf 95}, 236402 (2005)}


\begin{abstract}
  Low-temperature thermodynamic properties of strongly interacting
  Fermi liquids with fermion condensate are investigated.  We
  demonstrate that the spin susceptibility of these systems exhibits
  the Curie-Weiss law, and the entropy contains a
  temperature-independent term.  The excessive entropy is released at
  the superconducting transition, enhancing the specific heat jump
  $\Delta C$ and rendering it proportional to the effective Curie
  constant.  The theoretical results are favorably compared with the
  experimental data on the heavy fermion metal CeCoIn$_5$, as well as
  $^3$He films.
\end{abstract} 

\maketitle

Theoretical understanding of strongly interacting Fermi systems, such
as heavy-fermion metals, is challenging \cite{Coleman-2005}.
Conventionally, electrons in solids are classified as either itinerant
or localized.  The former have a Fermi surface, and their spin
susceptibility $\chi$ at low temperature $T$ follows the Pauli law
$\chi(T)=\rm const$, whereas the latter exhibit the Curie law
$\chi\propto1/T$.  The Curie law is observed in many heavy-fermion
metals \cite{Stewart-RMP-1984,Tayama-2002,Petrovic-2001,Sarrao-2002}
and is commonly attributed to the localized character of the $f$
electrons.  However, in some of these materials, such as CeCoIn$_5$
\cite{Petrovic-2001} and PuCoGa$_5$ \cite{Sarrao-2002,Curro-2005}, the
low-temperature Curie law is immediately followed by a superconducting
transition, also associated with the $f$ electrons \cite{Boulet-2005}.
If the $f$ electrons are localized, how can they superconduct?
Actually, measurements of the Fermi surfaces of the heavy fermion
metals by magnetic oscillations directly demonstrate that the $f$
electrons are itinerant, in agreement with band structure calculations
\cite{Lonzarich-1988}.  However, if the $f$ electrons are itinerant,
how can they exhibit the low-temperature Curie law?

We show that these puzzles can be resolved within the Fermi-liquid
theory if itinerant electrons form the so-called ``fermion
condensate'' state \cite{KS-1990,Nozieres}.  The interplay between
band-like and atomic-like behavior of electrons in solids is often
treated on the basis of the Hubbard model and the dynamical mean-field
theory \cite{Kotliar}.  Heavy fermions are typically described by the
Anderson-Kondo lattice models of coupled itinerant and localized
electrons originating from different orbitals, sometimes using the
two-fluid description \cite{Pines}.  However, given the experimental
evidence that the $f$ electrons are itinerant in some heavy-fermions
metals, here we study a conceptually simpler model where all electrons
are itinerant.  Our goal is not to present a detailed,
material-specific description, but to illustrate general ideas also
applicable to other puzzling Fermi systems, such as $^3$He films
\cite{Bauerle-1998,Casey-2003}.

Let us consider a system of itinerant electron quasiparticles
characterized by dispersion $\varepsilon_{\bm p}$, where $\varepsilon$
is energy measured from the chemical potential, and $\bm p$ is
momentum.  The spin susceptibility $\chi_0$ per one electron is
\begin{equation}
  \chi_0=-\mu_e^2\int{dn(\varepsilon_{\bm p})\over d\varepsilon_{\bm p}}
  \,d\upsilon_p  \equiv {\mu_e^2\over T}
  \int n({\bm p})\,[1-n({\bm p})]\,d\upsilon_p \ ,
\label{chi0}
\end{equation}
where $d\upsilon_p=2d^3p/\rho\,(2\pi\hbar)^3$ is the volume element in
3D momentum space, $\rho$ is the electron concentration, and $\mu_e$
is the magnetic moment of electron in solid.  In a simple case,
$\mu_e$ is equal to the Bohr magneton $\mu_B$, but, for heavy
fermions, $\mu_e$ may also contain a contribution from the orbital
angular momentum, as discussed later in the paper.  The occupation
numbers $n({\bm p})$ are given by the Fermi distribution function
\begin{equation} 
  n({\bm p})=[1+\exp(\varepsilon_{\bm p}/T)]^{-1}
\label{fst}
\end{equation} 
where the Boltzmann constant $k_B$ is set to 1.  In ordinary Fermi
liquids, $\varepsilon_p\approx v_F(p-p_F)$, where $p_F$ is the Fermi
momentum, and $v_F$ is the Fermi velocity.  In this case, the integral
in Eq.\ (\ref{chi0}) is proportional to $T$, and
$\chi_0(T)=\mu_e^2N_0/\rho=\rm const$, where $N_0=p_F^2/\pi^2 v_F$ is
the density of states at the Fermi level.  Accounting for the
spin-spin interaction amplitude $g_0$ modifies Eq.\ (\ref{chi0}) via
the Stoner factor: $\chi=\chi_0/(1-g_0\chi_0)$.

The quasiparticle dispersion $\varepsilon_{\bm p}$ is affected by the
Landau interaction function $f_L({\bm p},{\bm
  p}')=\delta\varepsilon_{\bm p}/\delta n({\bm p}')$.  In general,
$f_L({\bm p},{\bm p}')$ is a functional of the occupation numbers
$n({\bm p})$, but here, for the sake of illustration, we take
$f_L({\bm p},{\bm p}')$ as a given function.  Then, $\varepsilon_{\bm
  p}$ is related to the bare dispersion $\varepsilon^0_{\bm p}$ as
\begin{equation}
  \varepsilon_{\bm p}=\varepsilon^0_{\bm p} 
  + \int f_L({\bm p},{\bm p}')\,n({\bm p}')\,d\upsilon_{p'} \ .
\label{xi}
\end{equation}
The dispersion $\varepsilon_{\bm p}$ and the occupation numbers
$n({\bm p})$ are obtained by solving Eqs.\ (\ref{fst}) and (\ref{xi})
self-consistently.  When the interaction $f_L$ is weak, Eq.\ 
(\ref{xi}) merely renormalizes the Fermi velocity.  However, when the
interaction strength exceeds a critical value, the minimum of the
total energy at $T=0$ may be achieved in a radically different state
with the fermion-condensate \cite{KS-1990}.  In this state, the
quasiparticle spectrum is completely flat $\varepsilon_{\bm p}=0$ at
$T=0$ in some region of momentum space, where the occupation function
$n_*({\bm p})$ continuously interpolates between 0 and 1
\cite{KS-1990}.  The increase of kinetic energy in this state is
compensated by the decrease of interaction energy.  Nozi\`eres
\cite{Nozieres} demonstrated that, at low $T\neq0$, the momentum
occupation function remains the same $n({\bm p})=n_*({\bm p})$ in the
domain occupied by the fermion condensate.  Thus, the self-consistent
dispersion $\varepsilon_{\bm p}$ becomes temperature-dependent
\begin{equation}
  \varepsilon_{\bm p}=T\,\ln\left({1-n_*({\bm p}) 
  \over n_*({\bm p})}\right) \ , 
\label{xi-T}
\end{equation}
as follows from the inversion of Eq.\ (\ref{fst}).  The group velocity
$\partial\varepsilon_{\bm p}/\partial {\bm p}$ in Eq.\ (\ref{xi-T}) is
proportional to $T$, which generates a sharp peak in the density of
states with the height proportional to $1/T$ and results in unusual
thermodynamic properties discussed below.  Measurements of magnetic
oscillations in the heavy-fermion metals indeed show enormous
flattening of $\varepsilon_{\bm p}$ relative to the band-structure
calculations \cite{Lonzarich-1988}.

\begin{figure}[b]
  \includegraphics[width=0.75\linewidth]{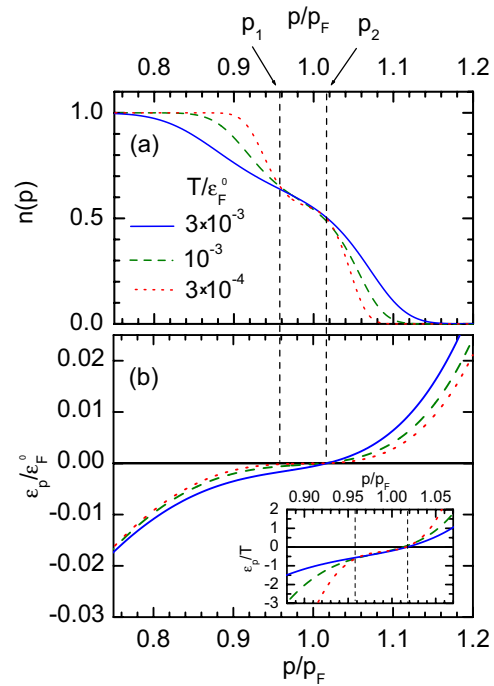}
\caption{Quasiparticle properties of a system with
  the fermion condensate, plotted vs.\ $p/p_F$ for three temperatures
  $T$: (a) the occupation numbers $n(p)$, (b) the single-particle
  spectrum $\varepsilon_p$ in the units of $\varepsilon_F^0=p_F^2/2m$,
  (inset) the ratio $\varepsilon_p/T$.}
\label{fig:fc}
\end{figure}

This qualitative analysis has been confirmed by analytical and
numerical solutions of Eqs.\ (\ref{fst}) and (\ref{xi}) for various
interaction functions $f_L$.  As an example, in Fig.\ \ref{fig:fc}, we
show $n(p)$ and $\varepsilon_p$ numerically calculated for three
different temperatures for a toy model with an isotropic parabolic
dispersion, characterized by the bare mass $m$ and the bare Fermi
energy $\varepsilon_F^0=p_F^2/2m$.  The interaction function $f_L(q)$,
where ${\bm q}={\bm p}-{\bm p}'$, was chosen to be
$f_L(q)=\lambda/\{[1-(q/2p_F)^2]^2+\beta^2\}$ with $\beta=0.48$ and
$\lambda=3p_Fm/\pi^2$.  Panel (a) shows that $n(p)$ markedly differs
from a step function and does not depend on temperature in the
interval $[p_1,p_2]$.  Panel (b) demonstrates that $\varepsilon_p$
changes with $T$, but the inset shows that the ratio $\varepsilon_p/T$
is $T$-independent in the interval $[p_1,p_2]$, in agreement with Eq.\ 
(\ref{xi-T}).  Fig.\ \ref{fig:fc} illustrates that a self-consistent
solution of quite conventional equations (\ref{fst}) and (\ref{xi})
for a rather generic, non-singular interaction function does generate
the fermion condensate.  Similar results were found for other
isotropic and crystal lattice models, where $\bm p$ is quasimomentum
in the Brillouin zone: see Ref.\ \cite{KCZ-condmat-2005} and
references therein.  The results are robust and do not depend
significantly on model details.  Although the toy model utilized for
the calculations shown in Fig.\ \ref{fig:fc} is not necessarily
realistic for specific materials, the fermion condensate formation is
a generic process, and only the fermion condensate parameters, not the
model details, matter for observable quantities.

\begin{figure}[b]
  \includegraphics[width=0.8\linewidth]{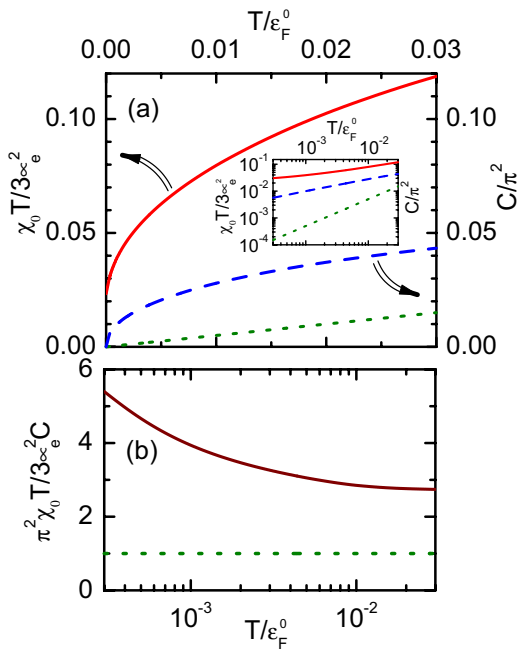}
  \caption{Thermodynamic properties of a
    fermion-condensate system. (a) The spin susceptibility
    $\chi_0T/3\mu_e^2$ (solid line) and the specific heat $C/\pi^2$
    (dashed line) vs.\ temperature $T$.  The inset: the same variables
    in the log-log scale.  (b) The Sommerfeld-Wilson ratio (solid
    line) vs.\ $T$.  The dotted lines in panels (a) and (b) correspond
    to the Fermi gas.}
\label{fig:chi-C}
\end{figure}

Now let us discuss observable manifestations of the fermion-condensate
state in detail.  Substituting the temperature-independent occupation
function $n_*({\bm p})$ of the fermion condensate into Eq.\ 
(\ref{chi0}), we find a Curie contribution to the spin susceptibility
\cite{ZK-JETP.Lett-2004}
\begin{equation}
  \chi_0={\kappa\,\mu_e^2 \over T}+\tilde\chi(T) \ , \quad 
  \kappa=\int_{p_1}^{p_2}n_*({\bm p})\,[1-n_*({\bm p})]\,d\upsilon_p \ ,
\label{Curie}  
\end{equation}
even though all electrons are itinerant.  The effective Curie constant
in Eq.\ (\ref{Curie}) is reduced by the dimensionless parameter
$\kappa$ relative to the Curie law $\chi_0=\mu_e^2/T$ of a
non-degenerate Fermi gas at high temperatures $T>\varepsilon_F^0$.
The second term $\tilde\chi$ in Eq.\ (\ref{Curie}) comes from
integration outside of the fermion-condensate domain $[p_1,p_2]$ in
Eq.\ (\ref{chi0}).  This term is less singular than the Curie term,
which dominates at low $T$.  Accounting for the spin interaction
amplitude $g_0$ generates the Curie-Weiss law
$\chi(T)\approx\mu_e^2\kappa/(T-\Theta_W)$ with the Weiss temperature
$\Theta_W=g_0\kappa\mu_e^2$.  The numerically calculated $\chi_0(T)$
for the same model as in Fig.\ \ref{fig:fc} is shown by the solid line
in Fig.\ \ref{fig:chi-C}a.  The finite value of the product $\chi_0T$
in the limit $T\to0$ indicates the Curie behavior at low temperatures
and gives the value $\kappa\approx0.1$ in this model.  In a wide
temperature range, $\chi_0(T)$ shown in Fig.\ \ref{fig:chi-C}a does
not strictly follow the Curie law because of $\tilde\chi(T)$.

In $^3$He films, the low-$T$ Curie constant is about 4 time lower than
the high-$T$ one, as shown in Fig.\ 1 of Ref.\ \cite{Bauerle-1998},
which gives $\kappa\approx0.25$ in this case.  To evaluate $\kappa$
for the heavy fermion metals from Eq.\ (\ref{Curie}), we need to know
the magnetic moment ${\bm \mu}$ of $f$ electrons, which has spin and
orbital contributions. In a free atom, ${\bm \mu}=g_L\mu_B{\bm J}$,
where $g_L$ is the Land\'e factor, and ${\bm J}$ is the total angular
momentum.  The crystal field lifts degeneracy between energy levels
with different projections $J_z$ and causes magnetic anisotropy.  In
CeCoIn$_5$, there is one $f$-electron with $J=5/2$, and the $\bm c$
axis is the easy magnetic axis.  Thus, the lowest energy levels have
$J_z=\pm5/2$, and the effective magnetic moment along the $\bm c$ axis
is $\mu_e=\mu_B g_L J_z=2.14\,\mu_B$, where $J_z=5/2$ and $g_L=0.857$
for $L=3$, $S=1/2$, and $J=5/2$.  As shown in Fig.\ 3 of Ref.\
\cite{Tayama-2002}, the low-$T$ Curie law in CeCoIn$_5$ is the most
pronounced for the easy axis $\bm c$ with the Curie constant
$0.2\mu_B^2$.  This value is much smaller than $\mu_e^2$, and we find
from Eq.\ (\ref{Curie}) that $\kappa=0.2\mu_B^2/(2.14\mu_B)^2=0.044$.

The entropy $S$ per one particle for an ensemble of fermion
quasiparticles is given by the formula
\begin{equation}
  S=-\int \{ n({\bm p})\ln n({\bm p}) + 
  [1-n({\bm p})]\ln[1-n({\bm p})] \}\, d\upsilon_p \ .
\label{entr}
\end{equation}
In ordinary Fermi liquids, the integrand in Eq.\ (\ref{entr}) differs
from zero only in a narrow vicinity of the Fermi surface, so
$S=T\pi^2/v_Fp_F$, and $S\to0$ when $T\to0$.  In contrast, in the
fermion-condensate state with the occupation function $n_*({\bm p})$,
the integrand is nonzero in a finite region, so the entropy has a
temperature-independent contribution $S_*$.  Extrapolating the
normal-state $S(T)$ from the inset in Fig.\ 2 of Ref.\ 
\cite{Petrovic-2001} to $T\to0$, we estimate that $S_*\approx0.1\ln2$
in CeCoIn$_5$.  We see that $S_*$ is finite, but much smaller than
$\ln2$ expected for an ensemble of localized spins 1/2.  Because it is
proportional to the momentum-space volume occupied by the fermion
condensate, $S_*$ may depend on external parameters, such as pressure
$P$, so that $\partial S_*/\partial P\neq0$.

In ordinary Fermi liquids, the specific heat $C=T(\partial S/\partial
T)$ is proportional to $T$.  The straight line of slope 1/2 in Fig.\
\ref{fig:chi-C}a shows $C/\pi^2$ and $\chi_0T/3\mu_e^2$ vs.\
$T/\varepsilon_F^0$ for a Fermi gas.  The fermion condensate does not
contribute to the specific heat, because its entropy $S_*$ is
$T$-independent.  So, the main contribution to $C$ comes from the
regions in momentum space where $\varepsilon_p$ interpolates between
the fermion condensate and the regular dispersion (see Fig.\
\ref{fig:fc}).  When the fermion-condensate domain is small,
$\varepsilon_p$ has inflection points
\cite{ZK-JETP.Lett-2004,CKZ-PRB-2005}, and the leading term in
$\varepsilon_p$ is proportional to $(p-p_1)^3$ for $\varepsilon_p<0$
and $(p-p_2)^3$ for $\varepsilon_p>0$, which gives $C\propto T^{1/3}$
and $\tilde\chi\propto T^{-2/3}$.  The dashed line in Fig.\
\ref{fig:chi-C}a shows the numerically calculated $C(T)$ for the same
model as in Fig.\ \ref{fig:fc}, which indeed exhibits a sublinear
power law.  Because the calculated $\chi_0T$ and $C$ have different
temperature dependences, the Sommerfeld-Wilson (SW) ratio
$R_{SW}^{(0)}=\pi^2\chi_0T/3C\mu_e^2$ increases with decreasing $T$,
as shown by the solid line in Fig.\ \ref{fig:chi-C}b and observed in
heavy fermions \cite{Gegenwart-2005}.  This is in contrast to ordinary
Fermi liquids, where $R_{SW}^{(0)}=1$, as shown by the horizontal line
in Fig.\ \ref{fig:chi-C}b.  Notice that the Stoner factor is not
included in our definition of $R_{SW}^{(0)}$.

Although the excessive entropy $S_*$ of the fermion condensate does
not contribute to the specific heat, it produces an enormous
enhancement of the thermal expansion coefficient $\alpha=\partial
V/\partial T\equiv -\partial S/\partial P$ and the Gr\"uneisen ratio
$\Gamma=\alpha/C$ \cite{ZKSB-1997}.  In ordinary Fermi liquids,
$S\propto T$, thus $\alpha\propto T$ vanishes at $T\to0$, and
$\Gamma(T)=\rm const$.  In contrast, for the fermion condensate, the
derivative $\partial S_*/\partial P$ is $T$-independent, so $\alpha$
has a finite value at $T\to0$.  Experiment \cite{Oeschler-2003} shows
that $\alpha$ is indeed temperature-independent at low $T$ and exceeds
typical values for ordinary metals by the factor of $10^3$--$10^4$.
With $\alpha\to\rm const$ and $C(T)\to0$, the Gr\"uneisen ratio
$\Gamma=\alpha/C$ diverges at low $T$, which is observed
experimentally \cite{Gruneisen-2003}.

However, the existence of the residual entropy $S_*$ at $T\to0$
contradicts the third law of thermodynamics (the Nernst theorem).  To
ensure that $S=0$ at $T=0$, localized spins order magnetically due to
spin-spin interaction.  Similarly, a system with the fermion
condensate must experience some sort of a low-temperature phase
transition eliminating the excessive entropy $S_*$.  Here we focus on
the second-order phase transition to a superconducting state
\cite{KS-1990}.  The progressive increase of the fermion-condensate
density of states with decreasing temperature facilitates
superconducting instability in one of the pairing channels: $s$, $p$,
$d$, etc.  Elementary excitations in a superconductor are the
Bogolyubov quasiparticles, whose spectrum $E_{\bm
  p}=\sqrt{\varepsilon_{\bm p}^2+\Delta_{\bm p}^2}$ has the energy gap
$\Delta_{\bm p}$.  The entropy of a superconductor is given by Eq.\ 
(\ref{entr}) with $n(\varepsilon_{\bm p})\to f(E_{\bm p})$, where
$f(E_{\bm p})$ are the occupation numbers of the Bogolyubov
quasiparticles.  Because of the energy gap, $f\to0$, and so $S\to0$ at
$T\to0$, thus the Nernst theorem is satisfied.  However, in order to
release the excessive entropy $S_*$, the specific heat jump $\Delta C$
at the transition temperature $T_c$ is enhanced.

The specific heat $C_s$ of a superconductor is $C_s=\int
d\upsilon_p\,E_{\bm p}\,df(E_{\bm p})/dT$. Taking the difference
between $C_s$ and $C_n$, the specific heat in the normal state, we
find the specific heat jump at $T_c$:
\begin{equation}
  \Delta C = C_s-C_n = -{1\over 2T_c} \int {d\Delta_{\bm p}^2\over dT}
  n({\bm p})\,[1-n({\bm p})]\,d\upsilon_p \ .
\label{dC}
\end{equation}
In the BCS theory, $d\Delta^2/2T_cdT=-4\pi^2/7\zeta(3)\approx-4.7$,
where $\zeta$ is the zeta function.  (For $d$-wave pairing, the number
is different, but the results do not change significantly.)  Comparing
Eqs.\ (\ref{chi0}) and (\ref{dC}), we find
\begin{equation}
 \Delta C = 4.7\, T_c\chi_0/\mu_e^2 \equiv 1.43\,C_n R_{SW}^{(0)} \ .
\label{rat2}
\end{equation} 
Eq.\ (\ref{rat2}) shows that the specific heat jump $\Delta C$ can be
expressed in terms of either $\chi_0(T_c)$ or $C_n(T_c)$.  For
ordinary Fermi liquids, where the Sommerfeld-Wilson ratio
$R_{SW}^{(0)}=1$, Eq.\ (\ref{rat2}) reproduces the familiar BCS
relation $\Delta C/C_n=1.43$.  However, in the fermion condensate
state, $R_{SW}^{(0)}$ changes with temperature, so $\Delta C/C_n$ does
not have a universal value.  Since $\chi_0\propto1/T$, the specific
heat jump (\ref{rat2}) is not proportional to $T_c$, but is related to
the fermion condensate parameter $\kappa$ in the Curie law
(\ref{Curie}) \cite{Shaginyan-2001}:
\begin{equation}
  \Delta C\approx4.7\,\kappa \ .
\label{dC-kappa}
\end{equation}  
Thus, the ratio $\Delta C/C_n$ can be very high when $T_c$ is low,
because $C_n\to0$ at $T\to0$ while $\Delta C$ is finite.

Let us apply this analysis to CeCoIn$_5$, where $T_c=2.3$ K.  In this
material \cite{Petrovic-2001}, $\Delta C/C_n\approx4.5$ is
substantially higher than the BCS value, in agreement with our
arguments.  Using the value $\kappa\approx0.044$ evaluated from the
Curie law, we estimate the r.h.s.\ of Eq.\ (\ref{dC-kappa}) as 0.21.
This is about a half of the experimental value of $\Delta
C\approx0.42$ per electron measured in Ref.\ \cite{Bauer-2004}.
However, as shown in Fig.\ 3 of Ref.\ \cite{Tayama-2002}, the Curie
term constitutes only about a half of the spin susceptibility at
$T_c$.  Thus, using the total susceptibility, we find that the
relation (\ref{rat2}) between $\Delta C$ and $\chi_0$ is satisfied.
In PuCoGa$_5$, we also attribute the Curie law, followed by a
superconducting transition at $T_c=18.5$ K \cite{Sarrao-2002}, to the
fermion condensate.  However, quantitative estimate of $\kappa$ is
difficult in this case, because plutonium has five $f$ electrons.
Interestingly, if plutonium is replaced by uranium, the resulting
material UCoGa$_5$ does not exhibit the Curie law and does not have
superconducting transition \cite{Sarrao-2002}.  This is surprising
from a conventional point of view, where ``itinerant'' electrons in
UCoGa$_5$ should be more susceptible to superconducting pairing than
``localized'' electrons in PuCoGa$_5$. However, if the fermion
condensate does not form in UCoGa$_5$, so that there is no Curie law,
then the density of states is not enhanced, and superconductivity is
not facilitated.

In conclusion, we have shown that strongly interacting Fermi liquids
can form a fermion-condensate state, where quasiparticle dispersion
$\varepsilon_{\bm p}$ is flat at the Fermi level.  Their magnetic
susceptibility $\chi_0(T)$ exhibits the Curie-Weiss law with the
effective Curie constant reduced by the fermion-condensate parameter
$\kappa$.  The entropy has the temperature-independent term $S_*$
(estimated as $S_*\approx0.1\ln2$ per electron in CeCoIn$_5$), which
greatly increases the thermal expansion coefficient $\alpha=-\partial
S/\partial P$ at low $T$.  The excessive entropy $S_*$ is released
below the superconducting transition temperature $T_c$, which
dramatically reduces $\alpha$ and enhances the specific heat jump
$\Delta C/C_n$, as observed in CeCoIn$_5$
\cite{Oeschler-2003,Petrovic-2001}.  The universal relation
(\ref{rat2}) between $\Delta C$ and $T_c\chi_0(T_c)$ can be tested
experimentally by checking whether the both quantities change
proportionally upon variation of external parameters, such as pressure
or chemical substitution \cite{Boulet-2005,Tanatar-2005}.

We thank A.~Chubukov, P.~Coleman, G.~Lonzarich, J.~Thompson,
J.~Saunders, F.~Steglich, and G. Stewart for discussions.  This work
was initiated at the Quantum Phase Transitions program an the Kavli
Institute for Theoretical Physics, Santa Barbara in 2005.


\end{document}